\documentclass[a4paper]{article}

\usepackage{INTERSPEECH2020}
\usepackage[]{subcaption}

\title{Incremental Text to Speech for Neural Sequence-to-Sequence Models using Reinforcement Learning}
\name{Devang S Ram Mohan$^1$, Raphael Lenain$^2$\textsuperscript{*}\thanks{\noindent\textsuperscript{*}work done while at Papercup Technologies Ltd.}, Lorenzo Foglianti$^1$, Tian Huey Teh$^1$, Marlene Staib$^1$, Alexandra Torresquintero$^1$, Jiameng Gao$^1$}

\address{
  $^1$Papercup Technologies Ltd.,
  $^2$ Novoic}
\email{devang@papercup.com}

\begin{document}

\maketitle
\begin{abstract}
    Modern approaches to text to speech require the entire input character sequence to be processed before any audio is synthesised. This latency limits the suitability of such models for time-sensitive tasks like simultaneous interpretation. Interleaving the action of reading a character with that of synthesising audio reduces this latency. However, the order of this sequence of interleaved actions varies across sentences, which raises the question of how the actions should be chosen. We propose a reinforcement learning based framework to train an agent to make this decision. We compare our performance against that of deterministic, rule-based systems. Our results demonstrate that our agent successfully balances the trade-off between the latency of audio generation and the quality of synthesised audio. More broadly, we show that neural sequence-to-sequence models can be adapted to run in an incremental manner.
\end{abstract}

\section{Introduction}

\label{sec:intro}

Efforts towards incremental text to speech (TTS) have typically focused on more traditional, non-neural architectures \cite{baumann2012inprotk, baumann-schlangen-2012-inpro, pouget2015hmm}. However, advancements in neural TTS \cite{shen2018natural, sotelo2017char2wav, vasquez2019melnet} have resulted in near human levels of naturalness and thus motivate an exploration of neural incremental TTS systems.

Neural TTS systems typically adopt sequence-to-sequence architectures which require the entire input sequence to be processed before generating any units of the output sequence. This offline characteristic is often useful; for example, a question mark at the end of a sentence would impact the intonation of preceding words. On the other hand, synthesising speech incrementally from text could be valuable. Such a model could be placed at the tail-end of an incremental speech recognition and machine translation pipeline to obtain a real-time speech-to-speech translation system.

\begin{figure}
    \centering
    \includegraphics[width=1.0\linewidth]{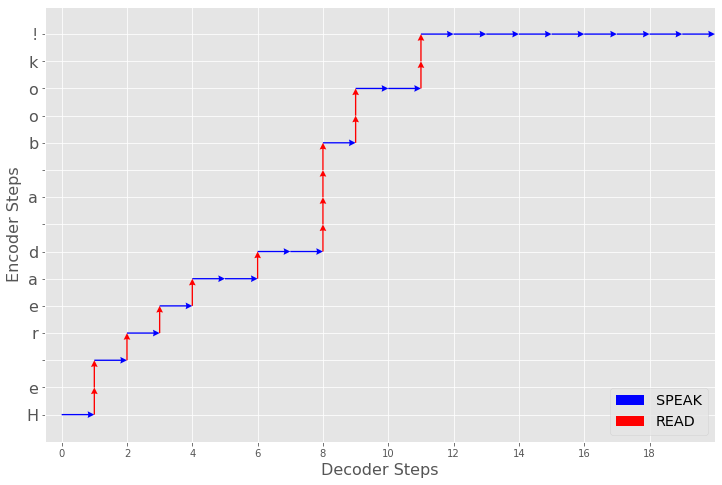}
    \caption{Trajectory for an arbitrary sequence of \textit{READ} (red, along y-axis) and \textit{SPEAK} actions (blue, along x-axis)}
    \label{fig:sample_policy}
\end{figure}

The development of these streaming, end-to-end architectures has seen considerable attention for the tasks of automatic speech recognition \cite{he2019streaming, zhang2020transformer, sak2017recurrent, jaitly2015neural} and machine translation (MT) \cite{DBLP:journals/corr/ChoE16, gu2016learning, zheng2019simpler, arivazhagan-etal-2019-monotonic}. Inspired by the approach of \cite{gu2016learning}, our proposed framework develops an agent that decides whether to trigger the encoder with the next input character (i.e., \textit{READ} in Figure \ref{fig:sample_policy}), or trigger the decoder with the characters read thus far (i.e., \textit{SPEAK} in Figure \ref{fig:sample_policy}). In this manner, our approach enables us to start generating mel-spectrograms while having read only a part of the input sentence. The mapping of these mel-spectrogram frames to raw audio waveforms can be achieved with an existing neural vocoder \cite{shen2018natural, DBLP:journals/corr/abs-1802-08435} by adjusting its inference behaviour.

The challenge then lies in deciding when to incorporate an additional character into this restricted input subsequence. We use the REINFORCE algorithm \cite{williams1992simple} to train an agent to make this decision.

\section{Background} \label{sec:background}

\cite{ma2019incremental} proposes an approach for incremental neural TTS. The model is based on the prefix-to-prefix framework \cite{ma2018stacl} and leverages a policy which maintains a fixed latency (in terms of number of words) behind the input. However, it would be challenging to construct such a rule-based approach if the desired latency was to be measured in a more granular unit, such as characters or phonemes. Furthermore, a dynamic, learnt policy would allow this approach to be used for new languages and speakers without manual calibration of these parameters.

The arena of incremental machine translation has also seen advancements. \cite{DBLP:journals/corr/ChoE16} proposes the framework of \textit{READ}/\textit{WRITE} and once again uses rule-based policies to enable incremental machine translation. \cite{gu2016learning} models this discrete action selection task using a reinforcement learning (RL) system, which we adapt in our work. Alternatively, \cite{zheng2019simpler} turns this non-differentiable framework into a supervised learning problem by training a model on sequences of interleaved \textit{READ}/\textit{WRITE} decisions generated from a pre-trained model.

A major challenge in any sequence transduction task is to align the target sequence with the source at each step. \cite{he2019streaming, zhang2020transformer} propose methods that leverage the RNN-T model \cite{graves2012sequence} to address this for the task of speech recognition. As an alternative, the approaches in \cite{sak2017recurrent, jaitly2015neural} propose architectures which utilise the fact that in speech recognition, the length of the target sequence is less than that of the source. \cite{DBLP:journals/corr/abs-1712-05382, DBLP:journals/corr/BahdanauCSBB15, tjandra-etal-2017-local} use encoder-decoder architectures with attention, but compute the attention alignments in an online manner.

\cite{1906.00672} adapts the online, monotonic attention mechanism proposed by \cite{raffel2017online} for the Tacotron 2 model. However, the motivation behind this was to ensure the surjectivity of the mapping between input elements and output frames and thus, the encoder and decoder architectures remain offline. Furthermore, the atomic input unit is a phoneme which can only be computed given the entire word. RL based approaches have also been used to generate attention weights for image captioning \cite{xu2015show}, \cite{DBLP:journals/corr/ZarembaS15, ling2017coarse}. However, these attention mechanisms generate hard attention weights which is undesirable for TTS \cite{battenberg2019locationrelative}.

\section{Tacotron 2 Modifications} \label{sec:tacotronmod}

Our base model builds on the Tacotron 2 model, with certain modifications for the incremental setting. Note that while these modifications may affect the quality of synthesised speech, they are necessary restrictions for incremental synthesis.

The encoder is altered by simply removing the convolutional layers and replacing the bi-directional LSTM \cite{schuster1997bidirectional, hochreiter1997long} with a uni-directional one. We further discard the post-net module, leaving only the attention mechanism that renders this model \textit{offline}. Rather than modifying the computation of the alignment weights and potentially enforcing a hardness constraint, we maintain the soft attention weights and suitably restrict its scope as described in Section \ref{sec:onlinettsusingrl}.

Finally, note that Tacotron 2 also has a vocoder component, which maps the mel-spectrogram to the raw audio waveform. We use a different vocoder architecture \cite{DBLP:journals/corr/abs-1802-08435} and adapt its inference behaviour to work in a purely auto-regressive manner by restricting the number of mel-spectrogram frames input to its residual and up-sampling networks.

For the remainder of this paper, we use this modified Tacotron 2 architecture to generate mel-spectrograms with the understanding that any incremental vocoder can be leveraged for synthesis.

\section{Incremental Text to Speech using Reinforcement Learning} \label{sec:onlinettsusingrl}

Inspired by \cite{gu2016learning}, we maintain an increasing buffer of input characters, which the model attends over to synthesise the next mel-spectrogram frame. We then train an agent to make the decision of whether to add the next input character into this buffer, or to synthesise a frame of audio based on the information in the buffer. To train this agent, we leverage the RL paradigm.

\subsection{RL Setup and Notation}

The RL setup consists of a decision maker, called the \textit{agent}, interacting with an \textit{environment}, typically over a sequence of discrete steps which we index by $j$. At the $j$th interaction step, the agent selects an action $a_{j}$, which the environment executes, and returns a new observation $\mathbf{o}_{j+1}$ (which is a representation of how its internal state has changed) and a numerical reward, $r_{j+1}$. In addition, the environment returns a flag which indicates whether this particular \textit{episode} of interactions has completed, called the terminal flag. The task for the agent, then, is to learn a mapping from the space of all possible observations to a suitable action. Such a mapping, called a \textit{policy}, should attempt to maximise the cumulative numerical reward achieved over the course of an episode (typically discounted temporally by a factor $\gamma \in [0, 1]$) \cite{sutton1998introduction}.

Formally, let $\mathbf{x}_{1}, ..., \mathbf{x}_{N}$ denote the sequence of input character embeddings and $\mathbf{h}_{1}, ..., \mathbf{h}_{N}$ denote the corresponding encoder outputs from our modified Tacotron 2 (Section \ref{sec:tacotronmod}). Our modifications enable $\mathbf{h}_{i}$ to be computed without knowledge of $\mathbf{x}_{i+1}, ..., \mathbf{x}_{N}$. Let the associated ground-truth mel-spectrogram $\mathbf{y} \in \mathbb{R}^{128 \times T}$ consist of $T$ frames. At the $j$th step of an episode, let $R(j) \in \{1, ..., N\}$ denote the number of characters that have been read and $S(j) \in \{1, ... T\}$ represent the number of audio frames generated (aligned by teacher-forcing \cite{shen2018natural} during training). Let $\alpha_{i, S(j)}$ denote the alignment weight over $\mathbf{h}_{i}$ while generating the $S(j)$th decoder output, $\mathbf{\hat{y}}_{S(j)}$. 

Instead of using \{$\mathbf{h}_{1}, ..., \mathbf{h}_{N}$\} to compute these weights (and thence the attention context), we use our restricted buffer $\{\mathbf{h}_{1}, ..., \mathbf{h}_{R(j)}\}$. This approach guarantees that, at the time of synthesising the $S(j)$th frame of audio, our Tacotron 2 model only has access to the first $R(j)$ characters.

\subsection{Agent} \label{subec:agentdesc}

The actions available to the agent are:
\begin{itemize}
    \item \textit{READ}: (step along the vertical axis in Figure \ref{fig:sample_policy}) Provides the attention mechanism with an additional character over which it may attend.
    \item \textit{SPEAK}: (step along the horizontal axis in Figure \ref{fig:sample_policy}) Results in the generation of a mel-spectrogram frame based on the characters read thus far. 
\end{itemize}

Then, a desirable learnt policy might be the agent learning to \textit{SPEAK} as soon as there is enough \textit{READ} context, and to resume \textit{READ}ing only when the existing context is fully synthesised. Observe that the offline behaviour can also be obtained as a specific policy (\textit{READ} all characters and then \textit{SPEAK} until all frames are synthesised).

\subsection{Environment} \label{subsec:environmentdesc}

The environment uses a trained modified Tacotron 2 model to provide the agent with the requisite information and feedback.

\subsubsection{Observations}

Suppose we have just received action $a_{j-1}$. The environment increments the appropriate counter ($R(j)$ or $S(j)$, based on $a_{j-1}$) and passes $\mathbf{h}_{1}, ..., \mathbf{h}_{R(j)}$ to the attention module, which computes $\alpha_{1, S(j)}, ..., \alpha_{R(j), S(j)}$. The context vector is then 
\begin{equation}
    \mathbf{c}_{S(j)} = \sum_{i = 1}^{R(j)}\alpha_{i, S(j)} \mathbf{h}_{i}
\end{equation}

Since we want $\mathbf{o}_{j}$ to contain enough information for the agent to decide whether to \textit{READ} or \textit{SPEAK}, we define $\mathbf{o}_{j}$ to be the concatenation of:
\begin{itemize}
    \item $\mathbf{c}_{S(j)}$: The attention context vector based on the $R(j)$ characters read thus far.
    \item $\mathbf{\alpha}_{.,S(j)}[k:]$: A fixed length moving window of the latest attention weights. This term was found to be crucial for learning a good policy.
    \item $\mathbf{y}_{S(j)}$ (during training) or $\mathbf{\hat{y}}_{S(j)}$ (during evaluation): The most recent mel-spectrogram frame.
\end{itemize}

\subsubsection{Rewards}

Underpinning our RL framework is the understanding that the quality of the generated output may trade-off against the delay incurred. Thus, we define our reward as
\begin{equation}
    r_{j} := r_{j}^{D} + r_{j}^{Q}
\end{equation}
where $r_{j}^{D}$ encourages low latency while $r_{j}^{Q}$ encourages high quality synthesis. Motivated by the treatment in \cite{gu2016learning}, we define 
\begin{equation}
    r_{j}^{D} := r_{j}^{CR} + r_{j}^{AP}
\end{equation}
where
\begin{itemize}
    \item $r_{j}^{CR}$ is a local signal to discourage consecutive \textit{READ} actions
    \begin{equation}
        r_{j}^{CR} := \omega \times (sgn(c_{j} - c^{*}) + 1)
    \end{equation} $c_{j}$ is a counter for consecutive \textit{READ}s, $c^{*}$ is an acceptable number of consecutive \textit{READ}s and $\omega < 0 $ is a hyper-parameter.
    \item $r_{J}^{AP}$ is a global penalty incurred only at the end of an episode
    \begin{equation}
        r_{J}^{AP} := \beta \times \lfloor d_{T} - d^{*} \rfloor_{+}
    \end{equation}
    Geometrically, $d_{T}$ corresponds to the average proportion of area under the policy path (Figure \ref{fig:sample_policy}). A value of $1$ for $d_{T}$ corresponds to \textit{READ}ing the entire input sequence before generating any output, while $0$ corresponds to the unattainable scenario of synthesising all the audio without \textit{READ}ing any characters. $d^{*}$ is a target value for $d_{T}$ and $\beta < 0 $ is a hyper-parameter.
\end{itemize}
Prior works in MT \cite{gu2016learning, ma2018stacl} have a detailed description of these terms. 

To compute $r_{j}^{Q}$, we use the mean squared error (MSE) between the ground truth and generated mel-spectrograms (aligned using teacher forcing). While the MSE is limited as a measure of perceived quality  \cite{DBLP:journals/corr/abs-1708-05987}, its usage as a training objective for our underlying Tacotron 2 model suggests it is suitable for our setting. We obtain a quality penalty term given by 
\begin{equation}
    r_{j}^{Q} := \lambda \times MSE(\mathbf{y}_{S(j)}, \mathbf{\hat{y}}_{S(j)})
\end{equation} where $\lambda < 0 $. When a \textit{READ} is executed, $r_{j}^{Q}$ is set to $0$.

\subsubsection{Terminal Flag}

At train time, there are two ways that the episode can terminate:
\begin{itemize}
    \item $R(j) = N$ (all the characters have been read) At this point the agent is forced to \textit{SPEAK} until $S(t) = T$. It is then given a cumulative reward for these \textit{SPEAK} actions.
    \item $S(j) = T$ (all the aligned mel-spectrograms have been consumed) At this point, the agent is given an additional penalty equal to the number of unread characters and the episode is terminated.
\end{itemize}

During inference, the episode runs until our Tacotron 2 model's termination criterion (i.e., the stop token) is triggered.

\subsection{Agent Setup and Learning}

The agent receives an observation $\mathbf{o}_{j}$ which is passed through a policy network consisting of a 512-dimensional GRU unit, a 2 layer dense network with ReLU non-linearity, and a softmax layer, to produce a 2-dimensional vector of action probabilities.

To learn these policy parameters $\theta$, we use the policy gradient method \cite{williams1992simple} which maximises expected cumulative discounted reward. However, as a variance reduction technique, we replace the discounted returns $G_{j}$ in the update, with a normalised advantage value \cite{mnih2014neural}. To compute this we subtract a baseline return, $b_{\phi}(\mathbf{o}_{j})$ (where $\phi$ parameterises a 3-layer fully connected network), and then normalise the result \cite{mnih2014neural, dsilverlectures}. To learn the baseline network parameters $\phi$, we minimise the expected squared loss between $G_{j}$ and $b_{\phi}(\mathbf{o}_{j})$. 

For both terms, the expectation is approximated by sampling a trajectory under the policy $\pi_{\theta}$. All parameters are trained jointly on collected batches of transitions.

\section{Experiments} \label{sec:experiments}

\subsection{Settings} \label{subsec:expsettings}

We use the LJ Speech dataset \cite{ljspeech17}, which consists of English audio from a single speaker. We partition this dataset into 12,000 train and 1,100 test/validation data points. We train our modified Tacotron 2 model for 300,000 iterations following the training routine in \cite{shen2018natural}.

We set the weights of each reward component, $\omega = -1$, $\beta = -10$ and $\lambda = -100$, to ensure that the scale of contribution is comparable. The target number of consecutive characters read, $c^{*}$ is set to $4$ while the target average proportion of area under the policy path, $d^{*}$ is set to $0.5$. These values are interpretable levers that allow the model's behaviour to be tweaked. The look-back of the attention window was set to $5$.

During training, actions are sampled according to the probabilities returned by the policy to encourage exploration of the observation space. While evaluating, actions are chosen greedily. We use a discount factor of $0.99$ and train on batches of collected transitions at the end of every $10$ episodes, using an Adam optimiser \cite{kingma2014adam} initialised with a learning rate of $10^{-4}$.

\subsection{Benchmark Policies} \label{subsec:benchmarks}

To gauge the performance of our agent, we used two types of benchmark policies, inspired by \cite{ma2019incremental, gu2016learning}:

\textbf{Wait-Until-End (WUE)}: \textit{Execute \textit{READ} actions until the text buffer is empty and then decode everything}. Since this policy has access to the entire input sentence at the time of decoding, this gives an upper bound on the quality of the synthesised speech, at the cost of the largest possible delay.

\textbf{Wait-k-Steps (WkS)}: \textit{Execute a \textit{READ} action every $k$ steps, and decode in between}. Despite incurring a smaller delay, the restricted access to the input sentence while decoding may impact the quality of the generated speech.

\subsection{Qualitative Analysis} \label{subsec:qualanalysis}

\begin{figure*}
    \centering
        \begin{subfigure}[b]{0.24\textwidth}
             \centering
                 \includegraphics[width=1.0\textwidth]{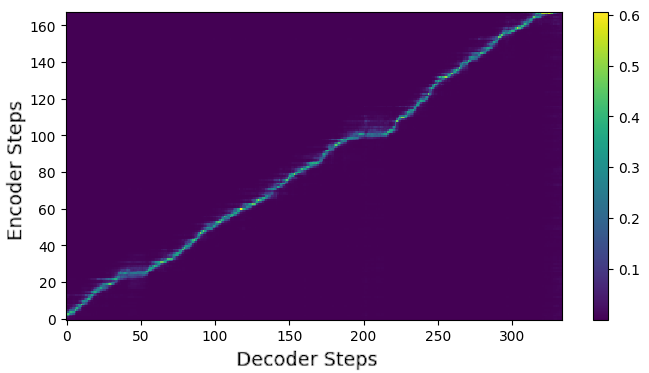}
             \caption{Wait Until End (WUE) Policy}
             \label{fig:read_every_1_alignments}
        \end{subfigure} %
        \begin{subfigure}[b]{0.24\textwidth}
             \centering
                 \includegraphics[width=1.0\textwidth]{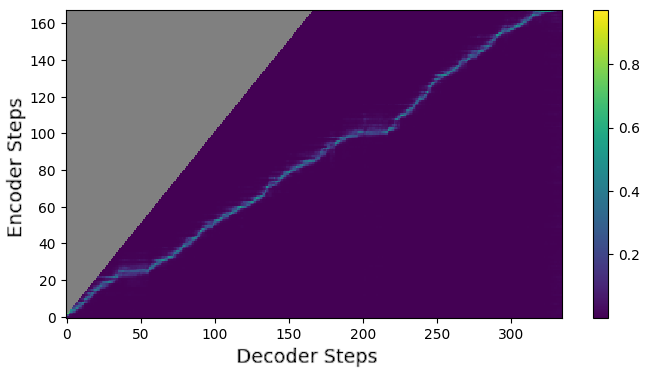}
             \caption{Wait 2 Steps (W2S) Policy}
             \label{fig:read_every_2_alignments}
        \end{subfigure} %
        \begin{subfigure}[b]{0.24\textwidth}
             \centering
                 \includegraphics[width=1.0\textwidth]{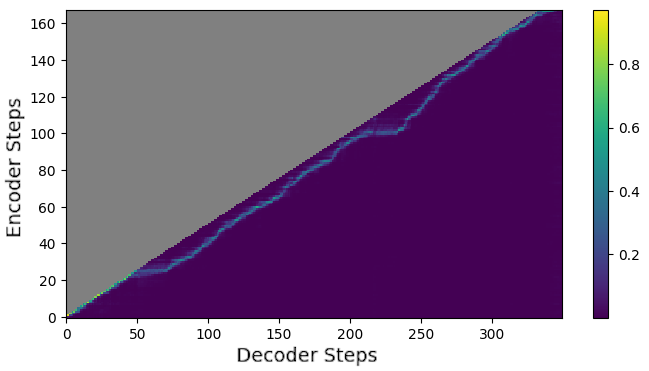}
             \caption{Wait 3 Steps (W3S) Policy}
             \label{fig:read_every_3_alignments}
        \end{subfigure} %
        \begin{subfigure}[b]{0.24\textwidth}
             \centering
                 \includegraphics[width=1.0\textwidth]{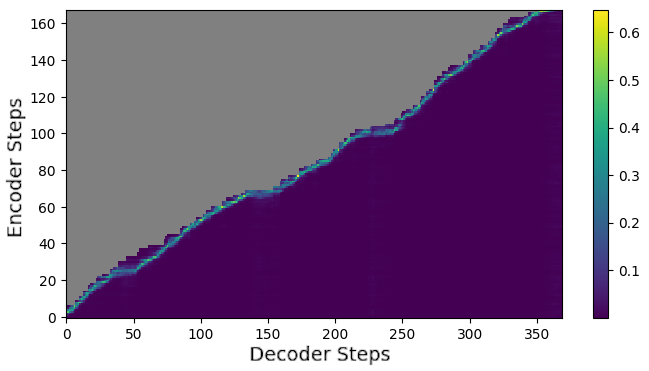}
             \caption{Learnt Policy}
             \label{fig:rl_policy_alignments}
        \end{subfigure}

    \caption[Different Policies with corresponding Attention Alignments]%
    {\textit{Policy Path with Attention Alignments (English)}: Each plot depicts the policy path and the attention alignments (by colour). The greyed out section represents portions of the input sentence that is unavailable as those input characters have not yet been read.}
    \label{fig:attention_plot_policy}
\end{figure*}

Figure \ref{fig:attention_plot_policy} depicts the attention alignments and policy path for a sample sentence \footnote{\label{foot:samples} English (and French) audio samples can be found at https://research.papercup.com/samples/incremental-text-to-speech}.
Figures \ref{fig:read_every_1_alignments} and \ref{fig:read_every_2_alignments} show that, for a large part of the decoding process, the WUE and W2S policies have access to more characters than required which highlights an avoidable latency. Figure \ref{fig:read_every_3_alignments} suggests that the W3S policy is able to reduce these unnecessary \textit{READ}s. However, the resulting policy path appears to collide with the `prominent' alignments on multiple occasions. As a result, the audio quality at these points is compromised because the decoder does not have sufficient context. This motivates the idea that an ideal policy path should \textit{hug} the prominent alignments diagonal closely from \textit{above} to successfully balance the quality of synthesis and latency incurred. Our learnt policy (Figure \ref{fig:rl_policy_alignments}) does precisely that. This suggests that the agent has in fact learnt to \textit{READ} only when necessary and \textit{SPEAK} only when it has something relevant to output.

\subsection{Quantitative Analysis} \label{subsec:quantanalysis}

There are two aspects of the agent's performance that we track:

\textbf{Quality:} We compute the Mean Opinion Score (MOS) to measure the naturalness of our audio \cite{streijl2016mean, shen2018natural}. We considered using a MUSHRA test \cite{series2014method}. However, since some policies may generate unintelligible samples of audio, which in turn could be scored below a noisy anchor, this approach was set aside. We are also interested in measuring the intelligibility of the synthesised speech. Automatic speech recognition systems use \textit{word-error rate} (WER) to measure the transcription quality \cite{ali2018word}. Following this approach, we obtain human transcriptions of the speech and compute the WER against the ground truth.

\textbf{Latency:} We use the proportion of area under the policy path, $d_{T} \in [0, 1]$ described in Section \ref{subsec:environmentdesc}. This metric lacks interpretability in terms of the actual delay incurred (e.g. the number of extra characters read). An alternate average lagging metric has been proposed in the MT setting \cite{ma2018stacl}. However, the skewed ratio between the source and target lengths for TTS coupled with a soft alignment between source and target make this metric challenging to adapt to TTS.

\subsubsection{Results}

\begin{figure}
    \centering
    \includegraphics[width=0.90\linewidth]{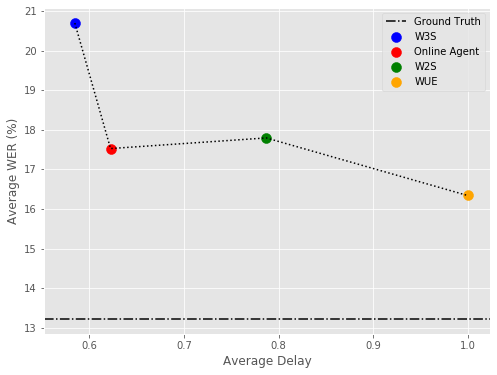}
    \caption[Average WER vs Average Proportion]{Average WER vs Latency ($d_{T}$) on a test set comprising 40 samples from LJ Speech labelled by 5 annotators}
    \label{fig:test_intelligibility}
\end{figure}

\begin{figure}
    \centering
    \includegraphics[width=0.90\linewidth]{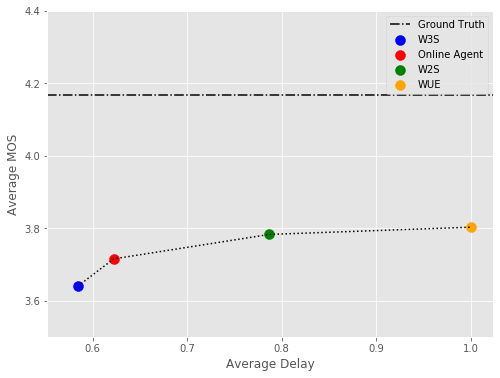}
    \caption[Average MOS vs Average Proportion]{Average MOS vs Latency ($d_{T}$) on a test set comprising 40 samples from LJ Speech labelled by 10 evaluators}
    \label{fig:test_naturalness}
\end{figure}

Figures \ref{fig:test_intelligibility} and \ref{fig:test_naturalness} depict the inherent trade-off between quality and latency. The ground truth marker depicts the value of the relevant metric for the vocoded ground truth mel-spectrograms. 

We begin by observing that the W3S policy incurs the least delay, closely followed by our online agent, while the W2S and WUE policies incur substantial delays. In terms of intelligibility, our online agent achieves a better WER than W3S, and even outperforms W2S despite its sizeable latency advantage. In terms of naturalness, our agent similarly outperforms W3S on MOS, but in this case, W2S was, as expected, able to leverage the additional latency to produce more natural sounding speech.

These findings establish that our agent is able to learn a policy that successfully balances the quality of the synthesised output against the latency incurred. The W2S policy is either comparable (intelligibility) or marginally better (naturalness) than our online agent, but in doing so, performs a large number of premature READ actions. Our agent incurs a slightly larger delay than the W3S policy, and manages to outperform it on all quality metrics. 

\section{Future Work}
\label{sec:futurework}

Our results show that for neural sequence-to-sequence, attention-based TTS models, there is no algorithmic barrier to incrementally synthesising speech from text. It is also interesting to analyse the learnt policy for different languages given the varied challenges posed (eg. elisions and liasons in French \cite{tranel1996french}). We provide samples from an agent trained on the French SIWIS dataset \cite{siwisdataset} with the same setup as described, on our samples page\footnotemark[1].

Furthermore, we used a modified Tacotron 2 model, pre-trained on full sentences. It would be interesting to analyse whether jointly learning the Tacotron weights helps synthesise partial fragments of a sentence better.

\section{Acknowledgements}
\label{sec:ack}
We would like to thank Simon King, Mark Herbster and Mark Gales for their valuable input on this research.

\bibliographystyle{IEEEtran}

\bibliography{mybib}

\begin{thebibliography}{10}
\providecommand{\url}[1]{#1}
\csname url@samestyle\endcsname
\providecommand{\newblock}{\relax}
\providecommand{\bibinfo}[2]{#2}
\providecommand{\BIBentrySTDinterwordspacing}{\spaceskip=0pt\relax}
\providecommand{\BIBentryALTinterwordstretchfactor}{4}
\providecommand{\BIBentryALTinterwordspacing}{\spaceskip=\fontdimen2\font plus
\BIBentryALTinterwordstretchfactor\fontdimen3\font minus
  \fontdimen4\font\relax}
\providecommand{\BIBforeignlanguage}[2]{{%
\expandafter\ifx\csname l@#1\endcsname\relax
\typeout{** WARNING: IEEEtran.bst: No hyphenation pattern has been}%
\typeout{** loaded for the language `#1'. Using the pattern for}%
\typeout{** the default language instead.}%
\else
\language=\csname l@#1\endcsname
\fi
#2}}
\providecommand{\BIBdecl}{\relax}
\BIBdecl

\bibitem{baumann2012inprotk}
T.~Baumann and D.~Schlangen, ``The inprotk 2012 release,'' in \emph{NAACL-HLT
  Workshop on Future Directions and Needs in the Spoken Dialog Community: Tools
  and Data}.\hskip 1em plus 0.5em minus 0.4em\relax Association for
  Computational Linguistics, 2012, pp. 29--32.

\bibitem{baumann-schlangen-2012-inpro}
\BIBentryALTinterwordspacing
------, ``{INPRO}{\_}i{SS}: A component for just-in-time incremental speech
  synthesis,'' in \emph{Proceedings of the {ACL} 2012 System
  Demonstrations}.\hskip 1em plus 0.5em minus 0.4em\relax Jeju Island, Korea:
  Association for Computational Linguistics, Jul. 2012, pp. 103--108. [Online].
  Available: \url{https://www.aclweb.org/anthology/P12-3018}
\BIBentrySTDinterwordspacing

\bibitem{pouget2015hmm}
M.~Pouget, T.~Hueber, G.~Bailly, and T.~Baumann, ``{HMM} training strategy for
  incremental speech synthesis,'' in \emph{Sixteenth Annual Conference of the
  International Speech Communication Association}, 2015.

\bibitem{shen2018natural}
J.~Shen, R.~Pang, R.~J. Weiss, M.~Schuster, N.~Jaitly, Z.~Yang, Z.~Chen,
  Y.~Zhang, Y.~Wang, R.~Skerrv-Ryan \emph{et~al.}, ``Natural {TTS} synthesis by
  conditioning {W}ave{N}et on mel spectrogram predictions,'' in
  \emph{ICASSP}.\hskip 1em plus 0.5em minus 0.4em\relax IEEE, 2018, pp.
  4779--4783.

\bibitem{sotelo2017char2wav}
J.~Sotelo, S.~Mehri, K.~Kumar, J.~F. Santos, K.~Kastner, A.~Courville, and
  Y.~Bengio, ``Char2wav: End-to-end speech synthesis,'' \emph{ICLR - Workshop
  Track}, 2017.

\bibitem{vasquez2019melnet}
S.~Vasquez and M.~Lewis, ``Melnet: A generative model for audio in the
  frequency domain,'' \emph{arXiv preprint arXiv:1906.01083}, 2019.

\bibitem{he2019streaming}
Y.~He, T.~N. Sainath, R.~Prabhavalkar, I.~McGraw, R.~Alvarez, D.~Zhao,
  D.~Rybach, A.~Kannan, Y.~Wu, R.~Pang \emph{et~al.}, ``Streaming end-to-end
  speech recognition for mobile devices,'' in \emph{ICASSP}.\hskip 1em plus
  0.5em minus 0.4em\relax IEEE, 2019, pp. 6381--6385.

\bibitem{zhang2020transformer}
Q.~Zhang, H.~Lu, H.~Sak, A.~Tripathi, E.~McDermott, S.~Koo, and S.~Kumar,
  ``Transformer transducer: A streamable speech recognition model with
  transformer encoders and {RNN-T} loss,'' \emph{arXiv preprint
  arXiv:2002.02562}, 2020.

\bibitem{sak2017recurrent}
H.~Sak, M.~Shannon, K.~Rao, and F.~Beaufays, ``Recurrent neural aligner: An
  encoder-decoder neural network model for sequence to sequence mapping.'' in
  \emph{INTERSPEECH}, 2017, pp. 1298--1302.

\bibitem{jaitly2015neural}
N.~Jaitly, D.~Sussillo, Q.~V. Le, O.~Vinyals, I.~Sutskever, and S.~Bengio, ``A
  neural transducer,'' \emph{arXiv preprint arXiv:1511.04868}, 2015.

\bibitem{DBLP:journals/corr/ChoE16}
\BIBentryALTinterwordspacing
K.~Cho and M.~Esipova, ``Can neural machine translation do simultaneous
  translation?'' \emph{arXiv preprint arXiv:1606.02012}, 2016. [Online].
  Available: \url{http://arxiv.org/abs/1606.02012}
\BIBentrySTDinterwordspacing

\bibitem{gu2016learning}
J.~Gu, G.~Neubig, K.~Cho, and V.~O. Li, ``Learning to translate in real-time
  with neural machine translation,'' \emph{arXiv preprint arXiv:1610.00388},
  2016.

\bibitem{zheng2019simpler}
B.~Zheng, R.~Zheng, M.~Ma, and L.~Huang, ``Simpler and faster learning of
  adaptive policies for simultaneous translation,'' \emph{Proceedings of the
  2019 Conference on Empirical Methods in Natural Language Processing}, pp.
  1349--1354, 2019.

\bibitem{arivazhagan-etal-2019-monotonic}
N.~Arivazhagan, C.~Cherry, W.~Macherey, C.-C. Chiu, S.~Yavuz, R.~Pang, W.~Li,
  and C.~Raffel, ``Monotonic infinite lookback attention for simultaneous
  machine translation,'' in \emph{ACL}, Jul. 2019, pp. 1313--1323.

\bibitem{DBLP:journals/corr/abs-1802-08435}
N.~Kalchbrenner, E.~Elsen, K.~Simonyan, S.~Noury, N.~Casagrande, E.~Lockhart,
  F.~Stimberg, A.~van~den Oord, S.~Dieleman, and K.~Kavukcuoglu, ``Efficient
  neural audio synthesis,'' \emph{PMLR}, vol.~80, pp. 2410--2419, 2018.

\bibitem{williams1992simple}
R.~J. Williams, ``Simple statistical gradient-following algorithms for
  connectionist reinforcement learning,'' \emph{Machine learning}, vol.~8, no.
  3-4, pp. 229--256, 1992.

\bibitem{ma2019incremental}
M.~Ma, B.~Zheng, K.~Liu, R.~Zheng, H.~Liu, K.~Peng, K.~Church, and L.~Huang,
  ``Incremental text-to-speech synthesis with prefix-to-prefix framework,''
  \emph{arXiv preprint arXiv:1911.02750}, 2019.

\bibitem{ma2018stacl}
M.~Ma, L.~Huang, H.~Xiong, K.~Liu, C.~Zhang, Z.~He, H.~Liu, X.~Li, and H.~Wang,
  ``{STACL}: Simultaneous translation with integrated anticipation and
  controllable latency,'' \emph{arXiv preprint arXiv:1810.08398}, 2018.

\bibitem{graves2012sequence}
A.~Graves, ``Sequence transduction with recurrent neural networks,''
  \emph{arXiv preprint arXiv:1211.3711}, 2012.

\bibitem{DBLP:journals/corr/abs-1712-05382}
C.~Chiu and C.~Raffel, ``Monotonic chunkwise attention,'' \emph{ICLR}, 2018.

\bibitem{DBLP:journals/corr/BahdanauCSBB15}
D.~Bahdanau, J.~Chorowski, D.~Serdyuk, P.~Brakel, and Y.~Bengio, ``End-to-end
  attention-based large vocabulary speech recognition,'' in
  \emph{ICASSP}.\hskip 1em plus 0.5em minus 0.4em\relax IEEE, 2016, pp.
  4945--4949.

\bibitem{tjandra-etal-2017-local}
\BIBentryALTinterwordspacing
A.~Tjandra, S.~Sakti, and S.~Nakamura, ``Local monotonic attention mechanism
  for end-to-end speech and language processing,'' in \emph{Proceedings of the
  Eighth International Joint Conference on Natural Language Processing (Volume
  1: Long Papers)}.\hskip 1em plus 0.5em minus 0.4em\relax Taipei, Taiwan:
  Asian Federation of Natural Language Processing, Nov. 2017, pp. 431--440.
  [Online]. Available: \url{https://www.aclweb.org/anthology/I17-1044}
\BIBentrySTDinterwordspacing

\bibitem{1906.00672}
M.~He, Y.~Deng, and L.~He, ``Robust sequence-to-sequence acoustic modeling with
  stepwise monotonic attention for neural {TTS},'' in \emph{INTERSPEECH}, 2019,
  pp. 1293--1297.

\bibitem{raffel2017online}
C.~Raffel, M.-T. Luong, P.~J. Liu, R.~J. Weiss, and D.~Eck, ``Online and
  linear-time attention by enforcing monotonic alignments,'' in
  \emph{Proceedings of the 34th International Conference on Machine
  Learning-Volume 70}.\hskip 1em plus 0.5em minus 0.4em\relax JMLR. org, 2017,
  pp. 2837--2846.

\bibitem{xu2015show}
K.~Xu, J.~Ba, R.~Kiros, K.~Cho, A.~Courville, R.~Salakhudinov, R.~Zemel, and
  Y.~Bengio, ``Show, attend and tell: Neural image caption generation with
  visual attention,'' in \emph{International conference on machine learning},
  2015, pp. 2048--2057.

\bibitem{DBLP:journals/corr/ZarembaS15}
W.~Zaremba and I.~Sutskever, ``Reinforcement learning neural turing machines,''
  \emph{arXiv preprint arXiv:1505.00521}, 2015.

\bibitem{ling2017coarse}
J.~Ling, ``Coarse-to-fine attention models for document summarization,'' Ph.D.
  dissertation, 2017.

\bibitem{battenberg2019locationrelative}
E.~Battenberg, R.~Skerry-Ryan, S.~Mariooryad, D.~Stanton, D.~Kao, M.~Shannon,
  and T.~Bagby, ``Location-relative attention mechanisms for robust long-form
  speech synthesis,'' 2019.

\bibitem{schuster1997bidirectional}
M.~Schuster and K.~K. Paliwal, ``Bidirectional recurrent neural networks,''
  \emph{IEEE Transactions on Signal Processing}, vol.~45, no.~11, pp.
  2673--2681, 1997.

\bibitem{hochreiter1997long}
S.~Hochreiter and J.~Schmidhuber, ``Long short-term memory,'' \emph{Neural
  computation}, vol.~9, no.~8, pp. 1735--1780, 1997.

\bibitem{sutton1998introduction}
R.~S. Sutton \emph{et~al.}, \emph{Introduction to reinforcement learning},
  2nd~ed., 1998.

\bibitem{DBLP:journals/corr/abs-1708-05987}
D.~Elbaz and M.~Zibulevsky, ``Perceptual audio loss function for deep
  learning,'' \emph{arXiv preprint arXiv:1708.05987}, 2017.

\bibitem{mnih2014neural}
A.~Mnih and K.~Gregor, ``Neural variational inference and learning in belief
  networks,'' \emph{ICML}, 2014.

\bibitem{dsilverlectures}
D.~Silver, ``Lectures on reinforcement learning,'' 2015.

\bibitem{ljspeech17}
K.~Ito, ``The {LJ} speech dataset,''
  \url{https://keithito.com/LJ-Speech-Dataset/}, 2017.

\bibitem{kingma2014adam}
D.~P. Kingma and J.~Ba, ``Adam: A method for stochastic optimization,''
  \emph{arXiv preprint arXiv:1412.6980}, 2014.

\bibitem{streijl2016mean}
R.~C. Streijl, S.~Winkler, and D.~S. Hands, ``Mean opinion score ({MOS})
  revisited: methods and applications, limitations and alternatives,''
  \emph{Multimedia Systems}, vol.~22, no.~2, pp. 213--227, 2016.

\bibitem{series2014method}
B.~Series, \emph{Method for the subjective assessment of intermediate quality
  level of audio systems}.\hskip 1em plus 0.5em minus 0.4em\relax Geneva:
  International Telecommunication Union, 2015.

\bibitem{ali2018word}
A.~Ali and S.~Renals, ``Word error rate estimation for speech recognition:
  e-{WER},'' in \emph{Proceedings of the 56th Annual Meeting of the Association
  for Computational Linguistics (Volume 2: Short Papers)}, 2018, pp. 20--24.

\bibitem{tranel1996french}
B.~Tranel, ``French liaison and elision revisited: A unified account within
  optimality theory,'' University of California, Irvine, 1996.

\bibitem{siwisdataset}
J.~Yamagishi, P.-E. Honnet, P.~Garner, and A.~Lazaridis, ``The {SIWIS} {F}rench
  speech synthesis database,''
  \url{https://datashare.is.ed.ac.uk/handle/10283/2353}, 2017.

\end{thebibliography}

\end{document}